\documentstyle[aps,prl,preprint,floats,epsfig]{revtex}

\begin{document}

\preprint{\tighten\vbox{\hbox{\hfil CLNS 97/1531}
                        \hbox{\hfil CLEO 97-30}
}}

\title{ Radiative Decay Modes of the $D^0$ Meson }

\author{CLEO Collaboration}
\date{\today}

\maketitle
\tighten

\begin{abstract} 
In this paper we describe a search for four radiative
decay modes of the $D^0$ meson: 
$D^0\to\phi\gamma$,
$D^0\to\omega\gamma$,
$D^0\to\bar{K}^{*}\gamma$, and
$D^0\to\rho^0\gamma$.
We obtain 90\% CL upper limits on the branching ratios
of these modes of
$1.9\times 10^{-4}$,
$2.4\times 10^{-4}$,
$7.6\times 10^{-4}$ and
$2.4\times 10^{-4}$
     respectively.
\end{abstract}
\newpage

{
\renewcommand{\thefootnote}{\fnsymbol{footnote}}

\begin{center}
D.~M.~Asner,$^{1}$ J.~Gronberg,$^{1}$ T.~S.~Hill,$^{1}$
D.~J.~Lange,$^{1}$ R.~J.~Morrison,$^{1}$ H.~N.~Nelson,$^{1}$
T.~K.~Nelson,$^{1}$ D.~Roberts,$^{1}$ A.~Ryd,$^{1}$
R.~Balest,$^{2}$
% ...
\end{center}
 
\small
\begin{center}
$^{1}${University of California, Santa Barbara, California 93106}\\
$^{2}${University of Colorado, Boulder, Colorado 80309-0390}\\
% ...
\end{center}
 

\setcounter{footnote}{0}
}
\newpage

\section{Introduction}                              
  Motivated by the successful CLEO II search for 
$ b \to s \gamma $ decays\cite{b2sg},
we have looked for analogous decays in the charm sector.
In this paper we consider decays of the pseudo-scalar $D^0$ 
meson to final states
consisting of a vector meson 
($\phi$,$\omega$,$\bar{K}^{*}$or $\rho$) plus a photon.

\begin{figure}
\psfig{file=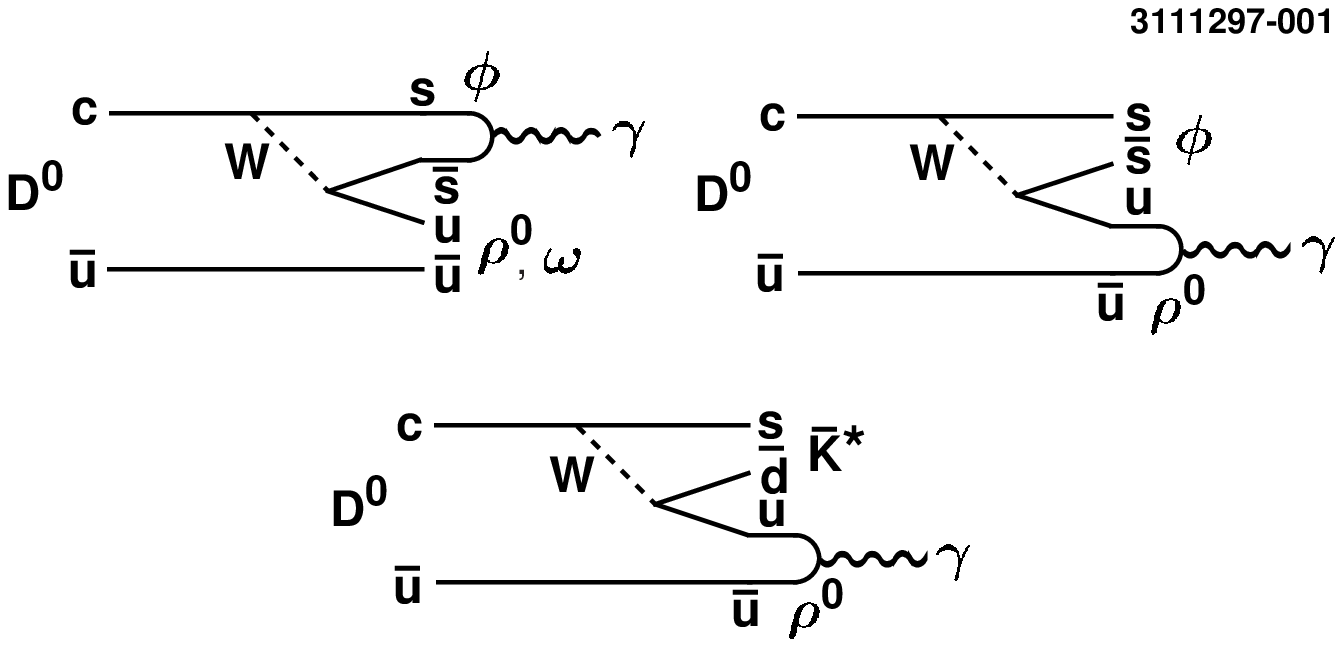}
\caption{}{\label{Fig:fyn}Feynman diagrams for the long distance 
electro-magnetic contributions.}
\end{figure}

Unlike $ b \to s \gamma $ decays, the short-range 
amplitudes relevant to $ c \to u \gamma $ are expected to 
be overwhelmed by much larger long-range electro-magnetic effects.
The dominant diagrams describing these electro-magnetic 
amplitudes are shown in Figure~\ref{Fig:fyn}.
In each case, a pair of vector mesons is produced.
Providing the quantum numbers are correct,  one of these can
couple to a photon.  The phenomenology of such interactions,
called ``Vector Meson Dominance'' (VMD),  has been well
studied \cite{vmd}.  Using VMD, one can make rough estimates of
the expected rates for the modes studied in this paper.  
If the coupling of the photon to the transverse component of a $\rho^0$
results in a vector conversion with about 1\% probability, we can use  the
Particle Data Group~\cite{pdg} value for the $ D^0 \to \phi \rho $
 branching ratio, $(2.6 \pm 0.8)\times 10^{-3}$,
and  expect that BR($D^0 \to \phi \gamma$) is about
$2.6\times 10^{-5} \cdot f_T$, where $f_T$ is the fraction of
$\rho$'s produced in the decay of the $D^0$  which are transversely polarized.
Detailed calculations of the long-range, $W$-exchange
and other contributing processes have been published 
by several groups \cite{pakvasa,bajc1,bajc2,cheng,fajfer}. The predictions range
from $10^{-4}$ to $10^{-6}$ and are listed in Table~\ref{Tbl:yield}.

  In the $b$ sector, observation of the decay
$B \to K^* \gamma $ at the measured rate provided compelling
evidence for the existence of a ``penguin''
contribution to the $B$ mesons decay amplitude into this 
channel.
The analogous short range penguin diagrams for the radiative decay of $D^0$ 
mesons are expected to contribute at the level of 
\mbox{${\cal B}_{c \to u \gamma}=10^{-11}-10^{-8}$} \cite{pakvasa,hurth},
making them relatively unimportant.

The long range electro-magnetic contributions that are expected
to dominate $D^0 \to V \gamma$ decay amplitudes also contribute
in the $b$ sector.  Their contribution to $B \to K^* \gamma $, for 
example, may be as big as 20\%\cite{golo}.
It is hoped that a study of these effects in the charm sector can 
improve our understanding of their relevance to bottom decay.

The CLEO collaboration has recently published a complimentary 
analysis searching for flavor changing neutral currents in 
$D^0\to Xl^+l^-$ decays\cite{fujino}.
  
\section{ Dataset and Event Selection}                              
  The data used for the analysis described in this
paper were acquired with
the CLEO II detector \cite{nim} at the Cornell
Electron Storage Ring (CESR), and represents a total 
integrated luminosity of $ 4.8 {\rm fb}^{-1}$.   

  When searching for $D^0 \to V \gamma$ decays we apply several selection
criteria on both the photon and vector meson candidates before attempting to
reconstruct the $D^0$\cite{cc}.  We look for $\phi \to K^+ K^-$ and
require \mbox{$1010 < M_{KK}({\rm MeV}/c^2) < 1030$}. 
We also demand that the time of flight 
and specific ionization of both $\phi$ daughter tracks be 
consistent with Kaon hypotheses. 
We require $\omega$ candidates to decay
into $\pi^+\pi^-\pi^0$ and  have $ 763 < M_{3\pi}({\rm MeV}/c^2) < 801 $.
Both photons from the $\pi^0$ are required to be in the central
region of the detector, $ | {\rm cos}(\theta_\gamma) | < 0.71 $, 
and  the $\gamma\gamma$ invariant mass must
be consistent with a $\pi^0$ with $\chi^2 < 4.8$.
To improve the measurement of the $\pi^0$ 4-vector, the photons 
are kinematically fit to the known $\pi^0$ mass.

We look for $\bar{K}^{*0} \to K^- \pi^+$ and
require \mbox{$842 < M_{K\pi}({\rm MeV}/c^2) < 942$}. 
In this mode we also make a cut on the decay angle of the
daughter particles in the $\bar{K}^{*}$ rest frame, requiring 
\mbox{$| {\rm cos} (\theta_{v\gamma}) | < 0.8$}, since signal events
should follow a ${\rm sin}^2(\theta_{v\gamma})$ distribution due to angular momentum
conservation.  Finally,
we reconstruct $\rho$'s through the decay $\rho \to \pi^+ \pi^-$ 
and require  $ 620<M_{\pi\pi}({\rm MeV}/c^2)<920 $.

In all cases we require that the ``radiative'' photon be in the 
central region of the calorimeter, have an energy greater 
than 830 MeV, and 
have a calorimeter shower isolated from charged
tracks in the event. To avoid background from $\pi^0$ decays
we veto photons that are part of a $\pi^0$ candidate 
with $\chi^2 < 15.3$. 

In this analysis, all $D^0$ candidates are required to come
from a $D^{*+}\to D^0\pi^+$ decay.  The additional kinematic
constraint provided by the $D^*$ is used to significantly reduce 
the otherwise large combinatoric background.  We require the reconstructed 
mass difference between the $D^{*+}$ and the $D^0$, 
\mbox{$\Delta M =  M(D^{*+}) - M(D^0)$}, to be
between 144.3 MeV/$c^2$ and 146.5 MeV/$c^2$. To further
reduce the background, 
we demand that $X_{D^*} > 0.625 $,
where $X_{D^*}$ is defined to be the momentum of the candidate $D^*$ divided 
by the maximum possible $D^*$ momentum.

The specific values of the cuts discussed above were chosen after
a systematic study of $(signal)^2/(background)$ for each of the modes, 
using large samples of GEANT\cite{geant} based Monte-Carlo data
to model each specific signal as well as the background.

\section{Backgrounds}                              
To learn about possible sources of background for each of
the four decay modes, a large sample of Monte-Carlo
generated  $e^+e^-\to q\bar{q}$ events was
analyzed.   The predominant
background source found was real $D^{*+} \to D^0 \pi^+$
decays where the $D^0$ decayed in channels 
involving $\pi^0$'s, which in turn
decayed such that one of the photons had very little energy
and went undetected.  Since the $D^*$ decay in the above 
sequence is real,
backgrounds of this kind will result in a false signal that
peaks in the mass difference $(\Delta M)$ distribution. Additional peaking
in the $D^0$ mass spectrum will
depend on kinematics.

\begin{figure}
\psfig{file=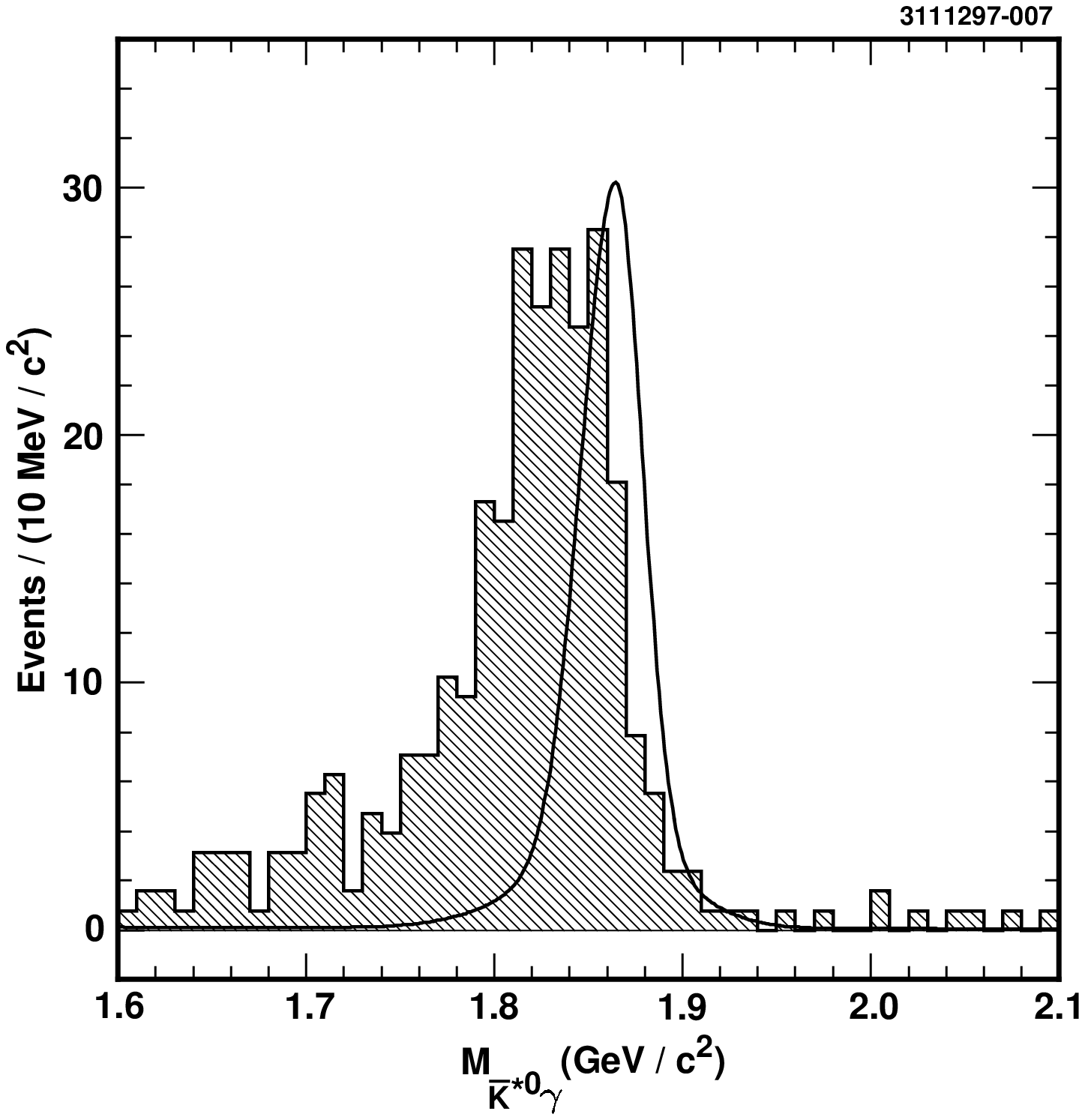,height=6in}
\caption{}{\label{Fig:kstback}
The correctly normalized background contribution from 
$D^0 \to K^- \pi^+ \pi^0$ Monte-Carlo events to the 
$D^0 \to \bar{K}^{*} \gamma$ invariant mass distribution (shaded histogram). 
The solid line shows the expected position and shape for real 
$D^0 \to \bar{K}^{*} \gamma$ events, also determined using Monte-Carlo.}
\end{figure}

This type of background is most severe for the 
$D^0 \to \bar{K}^{*} \gamma$ 
analysis because poorly reconstructed
$D^0 \to K^- \pi^+  \pi^0$ decays, where one of the
$\pi^0$'s photons is missed, will peak in the $D^0$ signal region.
Figure~\ref{Fig:kstback} shows the $D^0$
mass distribution for a set of $D^0 \to K^- \pi^+ \pi^0 $ events
analyzed as $ D^0 \to \bar{K}^{*} \gamma $.

\begin{figure}
\psfig{file=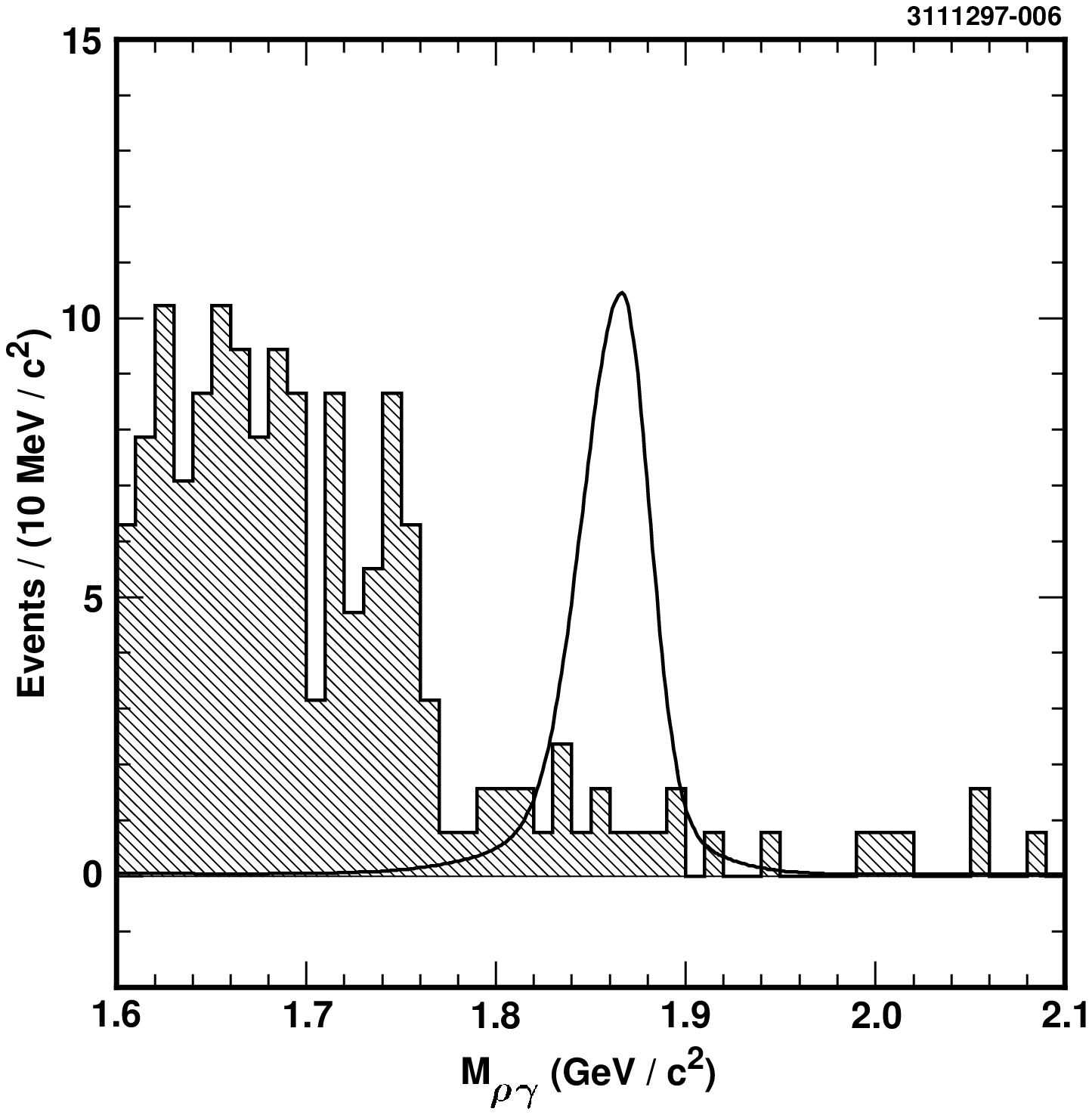,height=6in}
\caption{}{\label{Fig:rhoback}
The correctly normalized background contribution from 
$D^0 \to K^- \pi^+ \pi^0$ Monte-Carlo events to the 
$D^0 \to \rho \gamma$ invariant mass distribution (shaded histogram). 
The solid line shows the expected position and shape for real 
$D^0 \to \rho \gamma$ events, also determined using Monte-Carlo.}
\end{figure}

In the case of $D^0 \to \rho \gamma$ the problem is less severe since
there is no background decay mode which
peaks in the signal region of our invariant mass distribution, although
misreconstructed $D^0 \to K^- \pi^+ \pi^0 $ events cause the upward
distortion of the $D^0$ invariant mass spectrum just below the expected $D^0$ mass.
Figure~\ref{Fig:rhoback} shows the distribution of these
events when analyzed as $D^0 \to \rho \gamma$. 

For the modes $D^0 \to \omega \gamma $ and $ D^0 \to \phi \gamma $
there are no $D^0$ decay modes with large enough branching ratios 
to cause noticeable peaking in the reconstructed invariant mass
distribution, hence we expect the background in the $D^0$ mass spectra
of these to be smooth.

\section{Signal Yields and Limits}  
\begin{figure}
\psfig{file=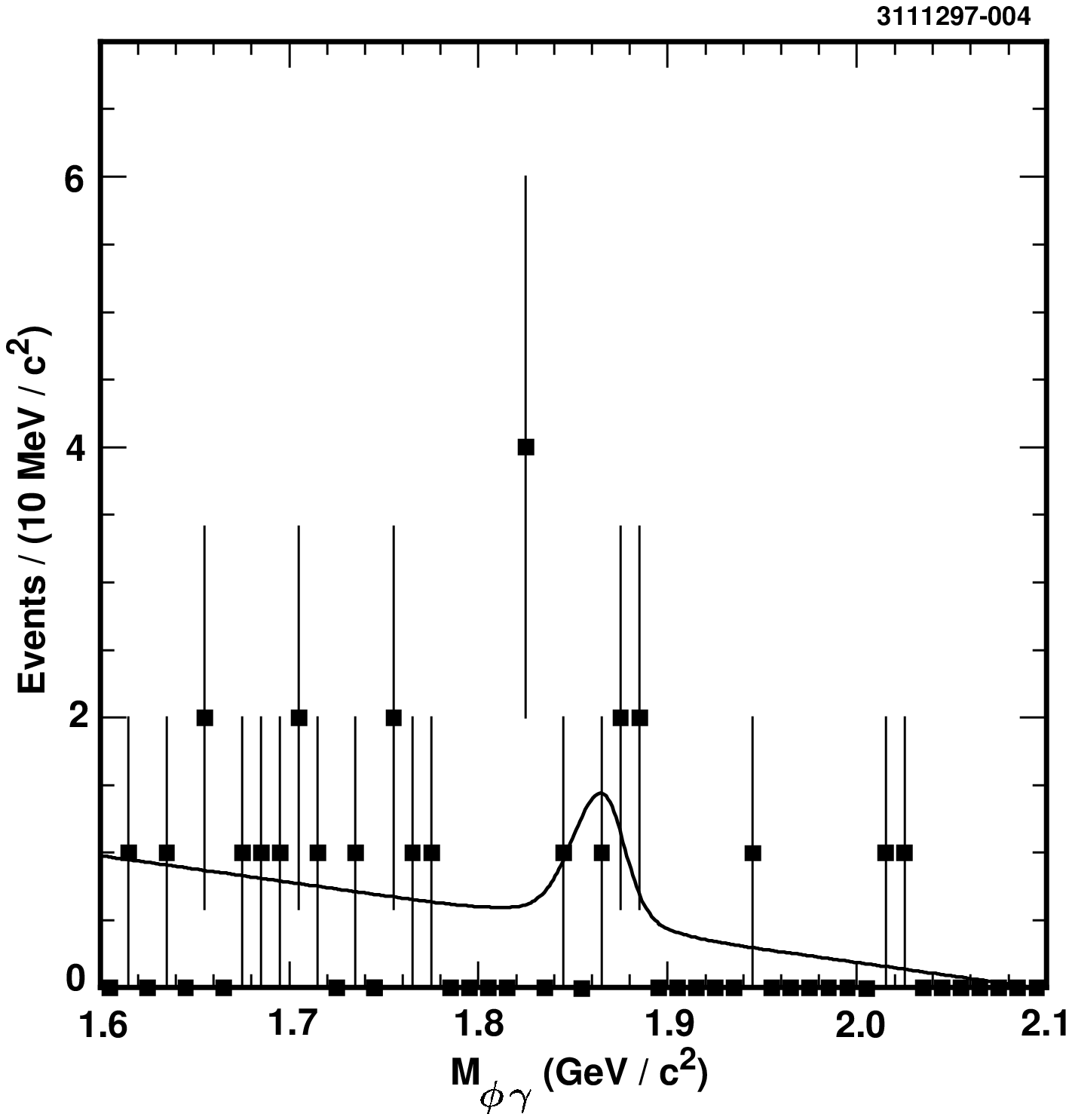,height=6in}
\caption{}{\label{Fig:phidat}Data and fit for the $D^0 \to \phi \gamma$ decay mode.}
\end{figure}

\begin{figure}
\psfig{file=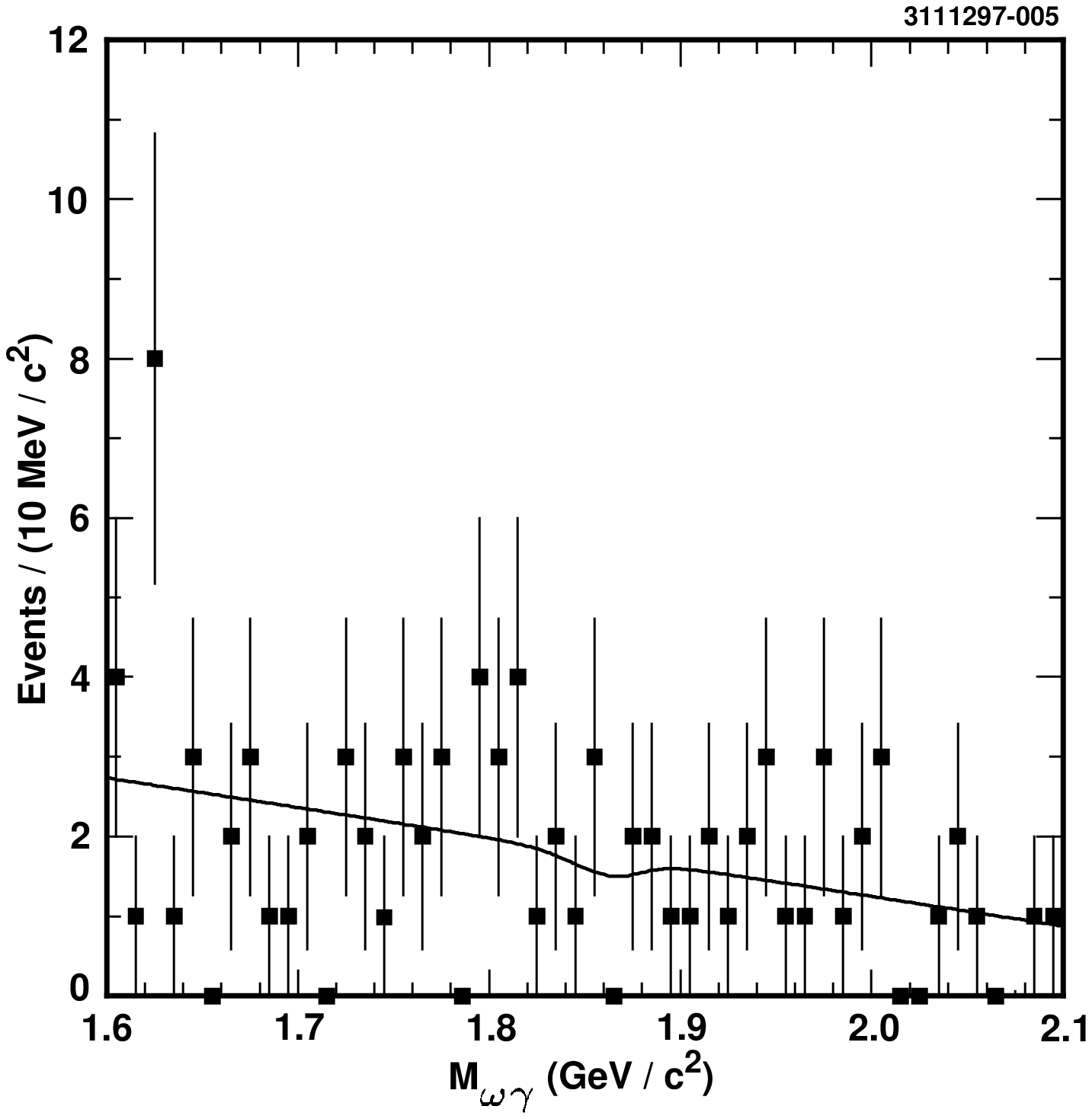,height=6in}
\caption{}{\label{Fig:omgdat}Data and fit for the $D^0 \to \omega \gamma$ decay mode.}
\end{figure}

  All yields were obtained by fitting the $D^0$ mass spectra. 
The signal in all cases was parameterized by a double
bifurcated  Gaussian whose mean and width were determined
using Monte-Carlo.  The background shape used depended on the
mode.
In the cases of $D^0 \to \phi \gamma$ and $D^0 \to \omega \gamma$
the background is expected to be smooth and likelihood fits were done
using simple linear background.  The data and fits for these modes are shown
in Figures~\ref{Fig:phidat} and~\ref{Fig:omgdat}.

\begin{figure}
\psfig{file=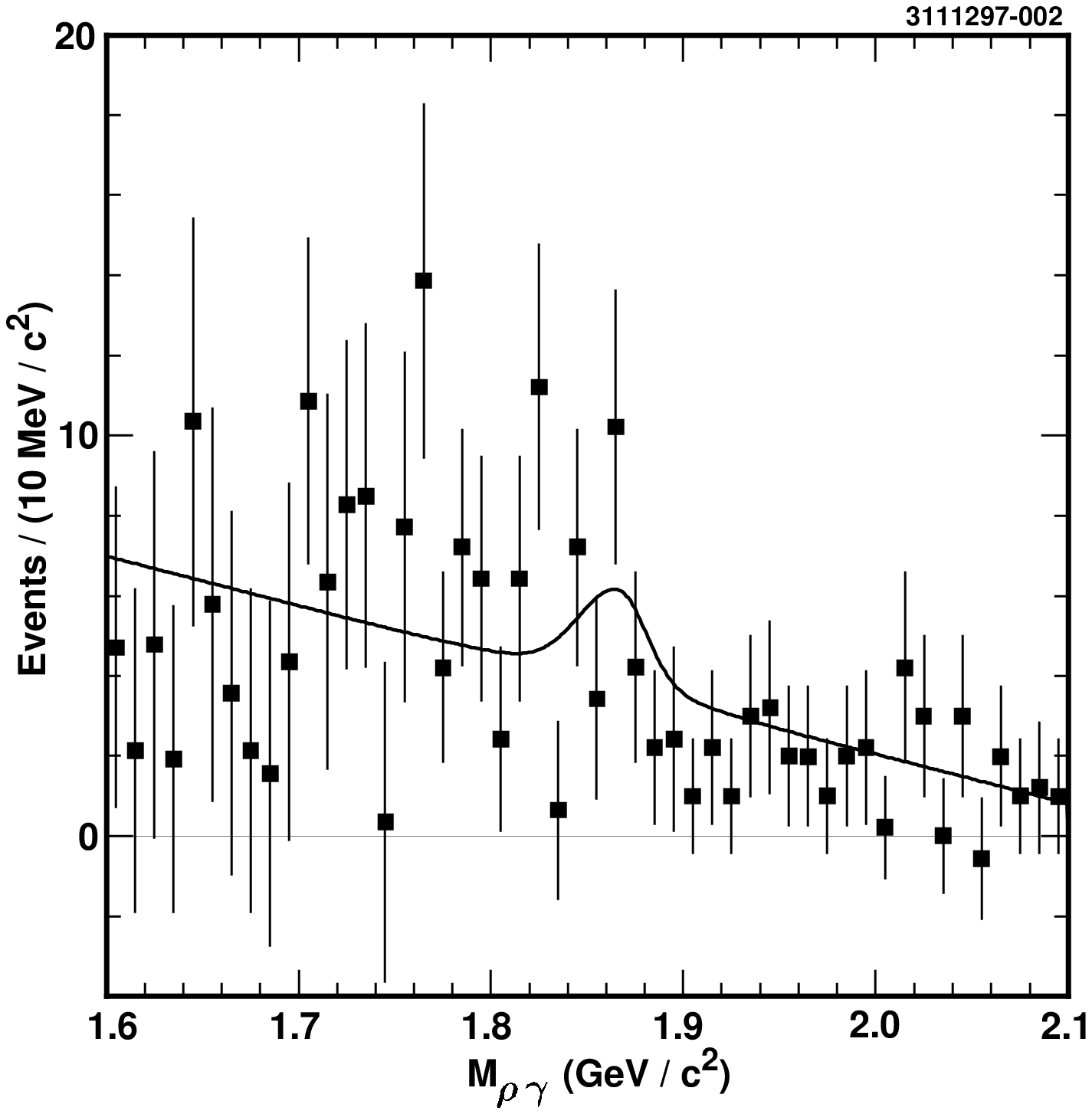,height=6in}
\caption{}{\label{Fig:rhodat}Data and fit for the $D^0 \to \rho \gamma$ decay mode. This plot shows the data
after subtraction of the $D^0 \to K^- \pi^+ \pi^0$ background estimation
from Monte-Carlo.}
\end{figure}

\begin{figure}
\psfig{file=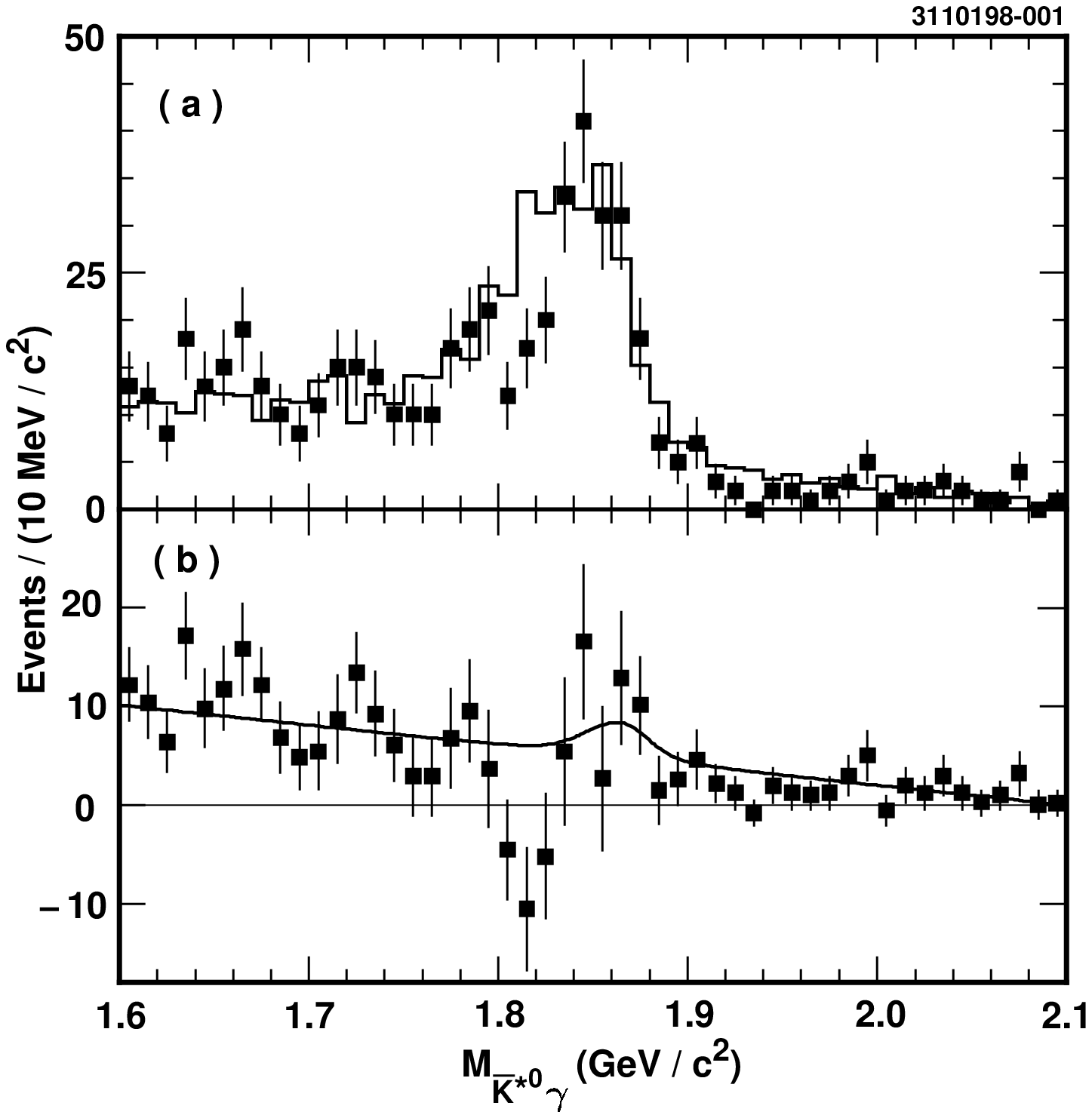}
\caption{}{\label{Fig:kstdat}
(a) The observed $D^0 \to \bar{K}^{*} \gamma$ data 
(points with error bars) and Monte-Carlo predicted background (solid histogram).

(b) Data and fit for the $D^0 \to \bar{K}^{*} \gamma$ 
decay mode after subtraction of the $D^0 \to K^- \pi^+ \pi^0$ background prediction.}
\end{figure}

  In the cases of $D^0 \to \rho \gamma$ and $D^0 \to \bar{K}^{*} \gamma$, we 
know the background shape is significantly modified by 
misreconstructed $D^0 \to {K}^{-} \pi^+ \pi^0$ decays.  
Using Monte-Carlo, we determined the magnitude and shape of this 
contribution to the $D^0$ invariant mass spectrum, and in both
cases included an additional component in our fits to compensate. 
The absolute normalization of this additional component was
determined from a previous analysis of 
$D^0 \to {K}^{-} \pi^+ \pi^0$ decays\cite{d2kpi}.
Figures~\ref{Fig:rhodat} and~\ref{Fig:kstdat}(b) show the 
mass spectra and  fits for
these modes after subtracting the contribution from 
misreconstructed $D^0 \to {K}^{-} \pi^+ \pi^0$.
  
The results are summarized in Table~\ref{Tbl:yield}. The efficiency 
for each mode was determined  by analyzing
samples of GEANT\cite{geant} based Monte-Carlo ``signal'' events,
and is also presented in Table~\ref{Tbl:yield}.
To obtain branching ratios from the efficiency corrected yields we
performed a parallel analysis looking for
$D^{*+} \to D^0 \pi^+$, $D^0 \to {K}^{-} \pi^+$ decays.
Our yield in this mode was $13,077 \pm 124$ events with an overall
analysis efficiency of ($16.9 \pm 0.2$) \%, determined using Monte-Carlo.  
Using the PDG value of ($3.86\pm 0.14$) \%
for the $D^0 \to {K}^{-} \pi^+$ branching ratio we 
find the initial number of $D^{*+} \to D^0 \pi^+$ decays in our 
data sample was $(2005 \pm 77)\times 10^3$.

\section{Systematic Errors}
Several sources of possible systematic error were investigated,
and the results are presented in Table~\ref{Tbl:sys}. 
With the exception of $D^0 \to \omega \gamma$, the uncertainty 
in each case is dominated by uncertainties in fitting.
To investigate this error we systematically changed  either the 
combinatorial background
shape, the normalization of the $D^0 \to {K}^{-} \pi^+ \pi^0$ background 
component (in the $\rho\gamma$ and $\bar{K}^{*}\gamma$ cases only),
the signal shape, and the number of bins used in the fits.  Constant, linear 
and quadratic background functions were tried. 
Signal shapes were parameterized by  Gaussian,  Double
Gaussian,  bifurcated  Gaussian and  the  double  bifurcated
Gaussian shapes.  In each case we took the  largest  variation as 
our estimate of  the  systematic error.

As an additional check we excluded the signal region and 
fit only the background, using simple event counting in the signal region
combined with Poisson statistics to obtain the upper limits. 
The result of this procedure for $D^0 \to \bar{K}^{*} \gamma$, the mode having 
the otherwise biggest fitting uncertainties, is shown in 
Figure~\ref{Fig:kstdat}(a).  In this case 
we fitted the $D^0 \to \bar{K}^{*} \gamma$ data with a 
linear combinatorial component plus the absolutely normalized 
Monte-Carlo predicted $D^0 \to {K}^{-} \pi^+ \pi^0$ background, 
excluding the region between 
$1.75 ~{\rm GeV}/c^2$ and $1.90 ~{\rm GeV}/c^2$ from the fit.
We then count data and predicted background events in the 
same region to obtain a net
yield of $-33\pm 24$.  Using a conservative yield of $0\pm 24$ events 
results in a 90\% CL upper limit yield of 39 events, consistent 
with the original fitted result.

The vector meson mass cuts were studied by varying them
to produce a 10\% change in efficiency and reanalyzing
both data and Monte-Carlo with the new values to estimate 
the systematic error.

To estimate the errors associated with analysis requirements
common to all of the studied modes (the $D^*-D$ mass difference and 
$D^*$ scaled momentum) while avoiding 
the problem of low statistics in the modes of interest, we used
numbers obtained in a previous measurement of 
$D^0 \to K^- \pi^+ \pi^0$ \cite{d2kpi}.

From a CLEO study of the decays $\eta \to \gamma\gamma$ 
and $\eta \to \pi^0\pi^0\pi^0$, we assign a 5.5\% systematic 
error for uncertainty in the overall $\pi^0$ finding efficiency and 
a 2.5\% uncertainty for each individual photon. The systematic
error  due  to  particle  identification  was  estimated  by
removing that cut entirely and noting the change. 

The systematic errors on the yield and the efficiency were
treated separately when calculating the final upper limits.
The efficiency and normalization errors were combined in quadrature 
and the efficiency was reduced by the resulting factor. The fitting
systematic errors were used to increase the 90\% CL upper limit yields.

\section{Conclusion}

Using data representing $4.8 ~{\rm fb}^{-1}$ of integrated luminosity
acquired by the CLEO II detector at the Cornell Electron Storage Ring,
we have conducted a search for radiative decay modes of the $D^0$ meson.  
The final results for the 90\% confidence level upper limit branching ratios
for the modes studied are:
$$ {\cal B}(D^0 \to \phi \gamma) < 1.9 \times 10^{-4}\ @\ 90\% CL $$
$$ {\cal B}(D^0 \to \omega \gamma) < 2.4 \times 10^{-4}\ @\ 90\% CL $$
$$ {\cal B}(D^0 \to \bar{K}^{*} \gamma) < 7.6 \times 10^{-4}\ @\ 90\% CL $$
$$ {\cal B}(D^0 \to \rho \gamma) < 2.4 \times 10^{-4}\ @\ 90\% CL $$

We  note  that  all  of  these values  are  well  above  the
theoretical expectations as shown in Table~\ref{Tbl:yield}.
We hope that with more data from CESR, KEK and PEP-II B-factories 
it will be possible to provide improved measurements in the future.

We gratefully acknowledge the effort of the CESR staff in providing us with
excellent luminosity and running conditions.
J.P.A., J.R.P., and I.P.J.S. thank                                           
the NYI program of the NSF, 
M.S. thanks the PFF program of the NSF,
G.E. thanks the Heisenberg Foundation, 
K.K.G., M.S., H.N.N., T.S., and H.Y. thank the
OJI program of DOE, 
J.R.P., K.H., M.S. and V.S. thank the A.P. Sloan Foundation,
M.S. thanks Research Corporation, 
and S.D. thanks the Swiss National Science Foundation 
for support.
This work was supported by the National Science Foundation, the
U.S. Department of Energy, and the Natural Sciences and Engineering Research 
Council of Canada.

\begin{table}
\caption{}{\label{Tbl:yield} The upper limit yields extracted from the 
likelihood fit and the resulting 90\% confidence level upper limits on the 
branching fractions incorporating systematic uncertainties 
in yield and efficiency determination.}
\begin{tabular}{ccccc}
  Mode  & $D^0 \to \phi \gamma$  &  $D^0 \to \omega \gamma $ & $D^0 \to \bar{K}^{*} \gamma$  & $D^0 \to \rho \gamma$ \\ \hline
  90\% CL Upper Limit Yield  &   8.9    &     7.7 &     38.5   & 21.6  \\
Detection Efficiency (\%)  & $ 5.57 \pm 0.13 \% $  & $ 2.10 \pm 0.05 \% $ & $ 5.51 \pm 0.13 \% $  & $ 5.83 \pm 0.13 \% $ \\ \hline
Branching Fraction \\
90\% CL Upper Limit  & $1.9 \times 10^{-4}$  & $2.4 \times 10^{-4} $ & $7.6 \times 10^{-4}$  & $2.4 \times 10^{-4}$ \\
Theoretical Prediction \cite{pakvasa,bajc1,bajc2,cheng,fajfer}  &  $ 0.01 - 0.34 \times 10^{-4} $  &  $ 0.01 - 0.09 \times 10^{-4} $ & $ 0.7 - 8.0 \times 10^{-4} $  & $ 0.01 - 0.63 \times 10^{-4} $ 
\end{tabular} 
\end{table} 

\begin{table}
\caption{}{\label{Tbl:sys}Estimated systematic errors for the four modes.}
\begin{tabular}{ccccc} 
  Mode  & $D^0 \to \phi \gamma$  &  $D^0 \to \omega \gamma $ & $D^0 \to \bar{K}^{*} \gamma$  & $D^0 \to \rho \gamma$ \\ \hline
Normalization  & 3.87\%  & 3.87\%  & 3.87\%  & 3.87\% \\
Monte-Carlo Stat.  & 2.25\%  & 2.4\% & 2.3\%  & 2.3\% \\
Branching Ratio of the Vector Meson  & 1.2\%  & 0.8\% & 0.1\%  & 0\%  \\
Photon and $\pi^0$ Eff.  & 2.5\%  & 8.5\% & 2.5\%  & 2.5\% \\
Vector-meson Mass Cut  & 3.4\%  & 2.6\% & 2.3\%  & 3.0\%  \\
Other Cuts   & 1.5\%  & 1.5\% & 1.5\%   & 1.5\% \\
 Particle  ID  &     9.2\%   &         ---   &       ---    &      --- \\ 
 Yield/Fitting    & 11.8\%  & 7.3\% & 38.8\%  & 23.6\% \\ \hline
Total Systematic Error  & 16\%  & 12\% & 39\%  & 24\% \\ \hline 
\end{tabular} 
\end{table}

.

\end{document}